\documentclass[letter,11pt]{article}
\usepackage{jheppub} % for details on the use of the package, please
\usepackage[T1]{fontenc} % if needed

\usepackage{slashed}
\usepackage{epsfig}
\usepackage{comment}
\usepackage{cancel}
\usepackage{bbm}
\usepackage{array}
\usepackage{bigints}
\usepackage{booktabs}
\usepackage{color}
\usepackage{dsfont}
\usepackage{float}
\usepackage{framed}
\usepackage{graphicx}
\usepackage{indentfirst}
\usepackage{mathrsfs}
\usepackage{multirow}
\usepackage{subdepth}
\usepackage{titlesec}
\usepackage[dotinlabels]{titletoc}
\usepackage{wrapfig}
\usepackage[all]{xy}
\usepackage[vcentermath]{youngtab}
\usepackage{relsize}
\usepackage{hyperref}
\numberwithin{equation}{section}

\usepackage[utf8]{inputenc}
\usepackage{slashed}
\usepackage{amsmath}
\usepackage{amsfonts}
\usepackage{amssymb}
\usepackage{mathtools}

\usepackage{cleveref}
\crefname{section}{§}{§§}
\Crefname{section}{§}{§§}

 \def\p{\partial}

\def\0{{(0)}}
\def\1{{(1)}}
\def\2{{(2)}}

\def\ci{{\mathscr I}}

\def\<{\langle }
\def\>{\rangle }

\def\l{{\lambda}}
\def\T{{\Theta}}

\def\G{{\Gamma}}
\def\cc{{\text{c.c.}}}
\def\o{{\omega}}

\newcommand{\bea}{\begin{eqnarray}}
\newcommand{\eea}{\end{eqnarray}}
\newcommand{\be}{\begin{equation}}
\newcommand{\ee}{\end{equation}}
\newcommand{\ba}{\begin{align}}
\newcommand{\ea}{\end{align}}

\newcommand{\w}[1]{\mbox{$\W_\infty[#1]$}}

   \makeatletter
  \let\over=\@@over \let\overwithdelims=\@@overwithdelims
  \let\atop=\@@atop \let\atopwithdelims=\@@atopwithdelims
  \let\above=\@@above \let\abovewithdelims=\@@abovewithdelims
\renewcommand\section{\@startsection {section}{1}{\z@}%
                                   {-3.5ex \@plus -1ex \@minus -.2ex}%nn
                                   {2.3ex \@plus.2ex}%
                                   {\normalfont\large\bfseries}}

\renewcommand\subsection{\@startsection{subsection}{2}{\z@}%
                                     {-3.25ex\@plus -1ex \@minus -.2ex}%
                                     {1.5ex \@plus .2ex}%
                                     {\normalfont\bfseries}}

\newcommand{\pd}[2]{\frac{\partial #1}{\partial #2}}

\newcommand{\beq}{\begin{equation}}
\newcommand{\eeq}{\end{equation}}
\newcommand{\beqa}{\begin{eqnarray}}
\newcommand{\eeqa}{\end{eqnarray}}
\newcommand{\beqar}{\begin{eqnarray*}}

\newcommand{\ve}{{\varepsilon}}

\def\YM{{\text{YM}}}

\def\[{\big[}
\def\]{\big]}

\def\adj{{\text{adj}}}

\def\ve{{\varepsilon}}

\def\g{{\gamma}}
\def\o{{\omega}}
\def\a{{\alpha}}

\def\w{\omega}
\def\be{{\bar \epsilon}}

\def\CA{{\mathcal A}}

\def\CD{{\mathcal D}}

\def\CI{{\mathcal I}}
\def\ci{{\mathscr I}}
\def\CJ{{\mathcal J}}

\def\CL{{\mathcal L}}

\def\CO{{\mathcal O}}

\def\CQ{{\mathcal Q}}

\def\CS{{\mathcal S}}
\def\CT{{\mathcal T}}

\newcommand{\bra}[1]{\langle\,   #1\,    |}
\newcommand{\ket}[1]{ |\,   #1 \,  \rangle}
\newcommand{\braket}[2]{\langle\,   #1 \, | \, #2 \, \rangle}
\newcommand{\tr}[1]{\text{tr} \left[ #1 \right]}
\def\mzz{{\mathbb Z}}

\def\mfg{{\mathfrak g}}
\def\pmm{{(\pm)}}

\def\+{{(+)}}
\def\-{{(-)}}
\def\0{{(0)}}
\def\1{{(1)}}
\def\2{{(2)}}
\def\3{{(3)}}

\def\mfg{{\mathfrak g}}

\title{\boldmath Asymptotic Symmetries in $(d+2)$-Dimensional Gauge Theories}

\author[]{Temple He$^1$}
\author[]{and Prahar Mitra$^2$}

\affiliation[]{$^1$Center for Quantum Mathematics and Physics (QMAP), Department of Physics, \\
University of California, Davis, 1 Shields Avenue, Davis, CA 95616, USA}\affiliation[]{$^2$School of Natural Sciences, Institute for Advanced Study, \\
1 Einstein Drive, Princeton, NJ 08540, USA}

\emailAdd{tmhe@ucdavis.edu}
\emailAdd{prahar21@ias.edu}

\abstract{We show that the subleading soft photon theorem in a $(d+2)$-dimensional massless abelian gauge theory gives rise to a Ward identity corresponding to divergent large gauge transformations acting on the celestial sphere at null infinity. We further generalize our analysis to $(d+2)$-dimensional non-abelian gauge theories and show that the leading and subleading soft gluon theorem give rise to Ward identities corresponding to asymptotic symmetries of the theory.}

\begin{document} 
\maketitle
\flushbottom

\section{Introduction}

The study of asymptotic symmetries in gauge and gravitational theories has recently experienced renewed interest due to the discovery that soft theorems in these theories are equivalent to Ward identities corresponding to the asymptotic symmetries \cite{Strominger:2013jfa}. This equivalence was originally studied in four-dimensional theories \cite{Campiglia:2014yka,Campiglia:2015kxa,Campiglia:2015qka,Campiglia:2015yka,Dumitrescu:2015fej,He:2014laa,Lysov:2015jrs,Mohd:2014oja,Campiglia:2016hvg,Conde:2016rom,He:2015zea,Kapec:2015ena,Kapec:2016jld,Mao:2017tey,Laddha:2017vfh,Hirai:2018ijc} and has since been generalized to all even spacetime dimensions \cite{Kapec:2014zla,Kapec:2015vwa}.

The generalization to odd spacetime dimensions, however, is complicated by the fact that waves propagate in qualitatively different manners in even and odd dimensions \cite{Balazs1955}. A related issue is that massless fields are non-analytic near $\ci^\pm$ in odd dimensions (which means one cannot perform a Taylor series expansion in $r^{-1}$), and it is not clear how to perform the asymptotic analysis in the presence of these non-analytic terms. These difficulties were ultimately surmounted in \cite{He:2019abc}, where the authors demonstrated that the leading soft photon theorem \cite{Weinberg:1965nx,Weinberg:1995mt} is equivalent to the Ward identity corresponding to large gauge symmetries in $U(1)$ gauge theory in \emph{any} dimension greater than or equal to four.

In addition to Weinberg's original leading soft theorems, there has also been progress in relating the subleading soft photon theorem to asymptotic symmetries of gauge theories in four dimensions \cite{Campiglia:2016hvg,Conde:2016csj,Laddha:2017vfh,Hirai:2018ijc}.\footnote{The results of \cite{Campiglia:2016hvg} and \cite{Conde:2016csj} are mathematically similar. In \cite{Campiglia:2016hvg}, the charges at $\ci^+_-$ are constructed by first taking $r\to\infty$ and then $u \to -\infty$ (which is what we do in this paper as well), whereas in \cite{Conde:2016csj}, the authors take $u,r\to\infty$ simultaneously while keeping $u\ll r$.} The corresponding symmetries are large gauge transformations that diverge linearly near null infinity. Given the results of \cite{He:2019abc}, it is natural to ask whether this equivalence can be extended to all dimensions greater than or equal to four.

We tackle this question in this paper and, more generally, attempt to present a systematic method that allows us to study the relationship between asymptotic symmetries and soft theorems in any massless gauge theory living in any dimension greater than or equal to four. In particular, we demonstrate that the subleading soft photon theorem corresponds to the Ward identity for divergent large gauge transformations in \emph{any} spacetime dimension. We further generalize the analysis to non-abelian gauge theories, thereby allowing us to relate both the leading and subleading soft gluon theorem to the corresponding Ward identities. 

Due to the fact that the present paper relies heavily on the techniques developed in \cite{He:2019abc}, we will refer the reader to that paper for various technical details. In an effort to make this paper self-contained, in \S\ref{sec:review}, we review \cite{He:2019abc} and present essential formulae that are necessary for our present work. In \S \ref{sec:subleading}, we study the subleading soft photon theorem and the corresponding Ward identity in a $U(1)$ gauge theory. Finally, in \S\ref{sec:nonabelian}, we generalize our results to non-abelian gauge theories and show that both the leading and subleading soft gluon theorem are associated to finite and divergent large gauge transformations, respectively. 

\section{Review of Previous Work}\label{sec:review}

In this section, we review the work of \cite{He:2019abc} and list the relevant notations, conventions, and formulae that will be utilized in the rest of this paper. We begin with $U(1)$ gauge theories in $(d+2)$-dimensions, which are described in terms of a field strength $F$ that satisfies Maxwell's equations
\begin{equation}
\begin{split}\label{maxwelleq}
\nabla^\mu F_{\mu\nu} = e^2 J_\nu , \qquad \p_{[\mu} F_{\nu\a]} = 0 ,
\end{split}
\end{equation}
where $J_\nu$ is a conserved matter current. The field strength can be separated into a radiative part $F^{(R)}$(which satisfies \eqref{maxwelleq} with $J=0$) and a Coulombic part $F^{(C)}$ (which is determined by a choice of Green's function). We will primarily be interested in the incoming and outgoing solutions (which correspond to the retarded and advanced Green's functions, respectively) and denote the corresponding radiative and Coulombic modes respectively by $F^{(R-)}$, $F^{(C-)}$ and $F^{(R+)}$, $F^{(C+)}$. 

\paragraph{Radiative Field} The radiative part admits a mode expansion, which in Cartesian coordinates $X^A$ takes the form
\begin{equation}
\begin{split} \label{gaugefieldmodeexp}
	F^{(R\pm)}_{AB} (X) = e \int \frac{d^{d+1} q}{ ( 2\pi )^{d+1} } \frac{1}{2q^0} \left[ \ve^a_{AB} ({\vec q}\,) \CO^\pmm_a(\vec{q}\,) e^{ i q \cdot X }  +  \ve^a_{AB} ({\vec q}\,)^* \CO^{\pmm\dag}_a(\vec{q}\,) e^{ - i q \cdot X }  \right] ,
\end{split}
\end{equation}
where $q^0 = | \vec{q}\,|$ and $\ve_{AB}^a(\vec{q}\,)$ are the $d$ polarization tensors labeled by $a$ and defined via 
\begin{equation}
\begin{split}\label{polvec}
	\ve^a_{AB}(\vec{q}\,) = i \left[ q_A \ve^a_B({\vec q}\,)  - q_B \ve^a_A({\vec q}\,)  \right] , \qquad q^A \ve_A^a ({\vec q}\,) = 0 , \qquad \eta^{AB}\ve^{a}_A({\vec q}\,) \ve_B^b({\vec q}\,)^* =  \delta^{ab} .
\end{split}
\end{equation}
In a quantum theory, the creation and annihilation modes satisfy the canonical commutation relation
\begin{equation}
\begin{split}
	\left[ \CO^{(\pm)}_a (\vec{q}\,) \,,\, \CO^{\pmm\dagger}_b (\vec{q}\,') \right] = \big( 2 q^0  \big) \delta_{ab} (2\pi)^{d+1} \delta^{(d+1)}  ( \vec{q} - \vec{q}\,'  ) .
\end{split}
\end{equation}

We are interested in the expansion of the field strength near $\ci^\pm$. To determine this, it is convenient to move to flat null coordinates, which are related to Cartesian coordinates via
\begin{align}\label{flatnullcoord}
X^A = \frac{r}{2}\left( 1+x^2 + \frac{u}{r},2x^{a},1-x^2 - \frac{u}{r}\right), \qquad x^2 = \delta_{ab}x^ax^b,
\end{align}
so that
\begin{equation}
\begin{split}
	ds^2 = \eta_{AB} dX^A dX^B = - du dr + r^2 dx^a dx_{a} . 
\end{split}
\end{equation}
In these coordinates, $\ci^\pm$ is located at $r \to \pm\infty$ keeping $(u,x)$ fixed. These surfaces have further boundaries at $u=\pm\infty$ that are denoted respectively by $\ci^+_\pm$ and $\ci^-_\pm$. The point labeled by $x^a$ on $\ci^+$ is antipodal to the point with the same label on $\ci^-$. We raise and lower lowercase Latin indices with $\delta^{ab}$ and $\delta_{ab}$, respectively.

A similarly convenient paramaterization is chosen for massless momenta:
\begin{equation}
\begin{split}\label{mompar}
	q^{A} = \o {\hat q}^A , \qquad {\hat q}^A = \left( \frac{ 1 + y^2}{2} \,,\,  y^a \,,\, \frac{ 1 - y^2 }{2} \right) , \qquad \ve^a_A ( {\vec q}\,) =  \p^a {\hat q}_{A}  =   \left( - y^a , \delta^a_b , - y^a  \right) . 
\end{split}
\end{equation}

We can now substitute \eqref{flatnullcoord} and \eqref{mompar} into \eqref{gaugefieldmodeexp}, determine the components of the field strength in flat null coordinates, and then take a large $|r|$ expansion. The expansion of the radial electric field in odd dimensions is ((2.23) in \cite{He:2019abc})
\begin{equation}
\begin{split}\label{Furlargerodd1}
F^{(R\pm)}_{ur} (u,r,x) &= \frac{e}{8(2\pi)^{\frac{d}{2}}  }   \sum_{n=0}^\infty   \sum_{s=0}^\infty  \frac{  2^{\nu_n - 2s} \csc ( \pi \nu_n )  }{ \G(s+1)\G ( 1 + s - \nu_n ) } \\
&\qquad \qquad \qquad \qquad \qquad  \times \left[ \frac{i (iu)^{s-\nu_n}  }{(ir)^{\frac{d}{2}+1+s}}  ( - \p^2 )^s  \p^{a} \CO^{(\pm,n)}_a ( x )   + \cc \right] \\
&\qquad \qquad \qquad + \frac{e}{2^{\frac{d}{2}+3}\pi^{d+1}  }   \sum_{n=0}^\infty \sum_{s=0}^\infty     (-1)^{s}   \frac{ \G\left( \frac{d}{2}+\nu_n+s \right)  }{ 2^{-\nu_n} \G(s+1)  }  \\
&\qquad \qquad \qquad  \qquad \qquad \times  \left[ \frac{i (iu)^{s} }{(ir)^{d+n+s}}   \int d^d y \frac{  \p^a \CO^{(\pm,n)}_{a} ( y )  }{\left[  ( x - y  )^2  \right]^{\frac{d}{2} + \nu_n + s }  }  + \cc \right]  , 
\end{split}
\end{equation}
whereas the analogous expansion in even dimensions takes the form ((2.27) in \cite{He:2019abc})
\begin{equation}
\begin{split}\label{Furlargereven1}
& F^{(R\pm)}_{ur} (u,r,x) \\
&= \frac{e}{2(2\pi)^{\frac{d}{2}+1}  }   \sum_{n=0}^\infty \sum_{s=0}^{\nu_n-1} \frac{ ( - 1 )^{s} \G(\nu_n-s)  }{ 2^{2s-\nu_n+1} \G(s+1) }  \left[ \frac{i ( iu)^{s-\nu_n}   }{(ir)^{\frac{d}{2}+1+s}}   ( - \p^2 )^s  \p^a \CO^{(\pm,n)}_{a} ( x )   + \cc \right] \\
&~ + \frac{e}{8\pi}   \sum_{n=0}^\infty \sum_{s=0}^\infty \frac{(-1)^{s}  2^{\frac{d}{2} + \nu_n} \G \left(\frac{d}{2}+s+\nu_n\right)   }{\G(s+1)  }  \left[ \frac{i (iu)^s    }{(ir)^{d+n+s}}   \int d^d y  \frac{\p^a \CO^{(\pm,n)}_{a} ( y ) }{  \left[(x-y)^2\right]^{\frac{d}{2}+\nu_n+s} }    + \cc \right]\\
&~ + \frac{e}{2(2\pi)^{\frac{d}{2}+1}  }   \sum_{n=0}^\infty \sum_{s=0}^\infty \frac{  2^{-2s-\nu_n} (-1)^{\nu_n-1} }{\G(s+1) \G(s+\nu_n+1) }   \\
&\qquad \times  \left[ \frac{i (iu)^{s}  \left[  \log \left(  \frac{1}{2} \frac{ \sqrt{iu} }{ \sqrt{ir}}  e^{\g_E} \right) - \frac{ 1 }{ 2  } \left( H_s + H_{s+\nu_n}  \right)  \right] }{(ir)^{d+n+s}}  ( - \p^2 )^{s+\nu_n} \p^a \CO^{(\pm,n)}_{a} ( x )  + \cc \right]  .   
\end{split}
\end{equation}
Here, we have defined $\nu_n = \frac{d}{2}-1+n$, $\gamma_E$ the Euler-Mascheroni constant, $H_n$ the $n$-th harmonic number (we define $H_0=0$), and $K_\nu$ the modified Bessel function of the second kind.

\paragraph{Coulombic Field} The asymptotic expansion of the Coulombic field is determined as follows. We assume that the Coulombic field strength and current admit a Taylor series expansion at large $|r|$ consistent with the asymptotic behavior described in \cite{He:2019abc}. Substituting these Taylor series into Maxwell's equations \eqref{maxwelleq}, we obtain equations order-by-order in large $|r|$ that can then be solved. The \emph{leading constraint equation} is
\begin{equation}
\begin{split}\label{cons1}
\p_u F^{(C\pm,d)}_{ur}  = \frac{e^2}{2} J^{(\pm,d)}_u , \\
\end{split}
\end{equation}
where throughout this paper $f^{(\pm,n)}$ denotes the coefficient of $|r|^{-n}$ in the expansion of the bulk field $f(u,r,x)$ near $r=\pm\infty$.

Furthermore, we will also require an additional constraint derived from the subleading terms in the large $|r|$ expansion of Maxwell's equations. These are obtained from the leading terms in the $r$ and $a$ components of Maxwell's equations and take the form
\begin{equation}
\begin{split}\label{subcons}
2 \p_u F^{(C\pm,d+1)}_{ur} - \p^a F^{(C\pm,d-1)}_{ua} &= e^2 J^{(\pm,d+1)}_u  \\
2 F_{ur}^{(C\pm,d+1)} \mp \p^a F_{ra}^{(C\pm,d)} &= \pm e^2 J_r^{(\pm,d+2)}  \\
- 2 \p_u F_{ra}^{(C\pm,d)} \pm  2 F_{ua}^{(C\pm,d-1)} &= e^2 J_a^{(\pm,d)}  \\
\p_u F_{ra}^{(C\pm,d)} + \p_a F_{ur}^{(C\pm,d)}  \pm (d-1) F_{ua}^{(C\pm,d-1)}  &= 0 . 
\end{split}
\end{equation}
It can be shown that equations \eqref{cons1} and \eqref{subcons} together imply the \emph{subleading constraint equation}
\begin{equation}
\begin{split}\label{cons2}
\p^2_u F_{ur}^{(C\pm,d+1)}  =  \mp \frac{e^2}{4d} \left[  \p^2 J^{(\pm,d)}_u + (d-1)   \p_u \p^a J_a^{(\pm,d)} - 2d \p^2_u J_r^{(\pm,d+2)}  \right] ,
\end{split}
\end{equation}
where $\p^2 \equiv \p^a \p_a$.

\section{Subleading Soft Photon Theorem}\label{sec:subleading}

\subsection{Matching Condition}

As discussed in \cite{He:2019abc}, the leading soft photon theorem is equivalent to a Ward identity derived by imposing the following antipodal matching condition across spatial infinity:
\begin{equation}
\begin{split}\label{oldmatchcond}
\left. F_{ur}^{(+,d)} \right|_{\ci^+_-}  &= -\left. F_{ur}^{(-,d)} \right|_{\ci^-_+} .\\
\end{split}
\end{equation}

The subleading soft photon theorem is derived from a similar matching condition for $F_{ur}^{(\pm,d+1)}$. There is, however, a slight obstruction to this. First, note that \eqref{oldmatchcond} is sensible only if $F_{ur}^{(\pm,d)}$ is finite and well-defined at $\ci^\pm_\mp$. This is indeed true, as one can verify from the expansions \eqref{Furlargerodd1} and \eqref{Furlargereven1} and the expressions for the Li\'enard-Wiechert electric field in $(d+2)$-dimensions. However, using \eqref{cons1} and \eqref{subcons} and assuming that the current has compact support in $u$, one finds that at large $|u|$
\begin{equation}
\begin{split}\label{udep}
F_{ur}^{(\pm,d+1)} = \mp \frac{1}{2d} u \p^2 F_{ur}^{(\pm,d)}  + F_{ur}^{(\pm,d+1)\text{fin}} ,
\end{split}
\end{equation}
where $F_{ur}^{(\pm,d+1)\text{fin}}$ is finite at large $|u|$. This structure can also be verified from the expansions \eqref{Furlargerodd1} and \eqref{Furlargereven1} of the radiative field strength. Thus, to impose a sensible matching condition, we match the finite part, i.e.
\begin{equation}
\begin{split}\label{finitematch}
\left. F_{ur}^{(+,d+1)\text{fin}}  \right|_{\ci^+_-}  &= \left.  F_{ur}^{(-,d+1)\text{fin}}   \right|_{\ci^-_+} .
\end{split}
\end{equation}
Noting that the projection operator $1-u\p_u$ projects out terms linear in $u$, we have\footnote{A similar projection was also used in \cite{Dumitrescu:2015fej} for the fermion operator.}
\begin{equation}
\begin{split}\label{finitematch1}
\left. F_{ur}^{(\pm,d+1)\text{fin}}  \right|_{\ci^\pm_\mp}   &= \left. ( 1 -  u\p_u ) F_{ur}^{(\pm,d+1)} \right|_{\ci^\pm_\mp} , 
\end{split}
\end{equation}
which implies
\begin{equation}
\begin{split}
\left. ( 1 -  u\p_u ) F_{ur}^{(+,d+1)} \right|_{\ci^+_-}  &= \left. ( 1 -  u\p_u )  F_{ur}^{(-,d+1)} \right|_{\ci^-_+} .
\end{split}
\end{equation}

We define the following charges
\begin{equation}
\begin{split}\label{matchcondsubleading}
\CQ^{\pm}_\lambda =  \frac{2}{e^2} \int_{\ci^\pm_\mp} d^d x\, \lambda ( 1 -  u\p_u ) F_{ur} ^{(\pm,d+1)}   , 
\end{split}
\end{equation}
where $\lambda \equiv \lambda(x)$ is a function defined on the celestial sphere at $\ci^\pm$. The antipodal matching condition immediately implies
\begin{equation}
\begin{split}\label{chargematch}
\CQ^+_\l = \CQ^-_\l.
\end{split}
\end{equation}
Using the covariant phase space formalism \cite{Wald:1999wa}, one finds that $\CQ_\lambda^\pm$ are related to the so-called \emph{divergent} large gauge transformations (see \cite{Campiglia:2016hvg,Laddha:2017vfh,Hirai:2018ijc} for a discussion of these symmetries). We can therefore think of these charges as measuring the local divergent $U(1)$ charge of the $in$ and $out$ states, and equivalently, \eqref{chargematch} is understood as a conservation law for these charges. Unlike the leading soft photon charge, this charge has no global counterpart since $\CQ^\pm_{\lambda=1} = 0$. Thus, even though there are local charges that are conserved due to the matching condition, there are no new global symmetries.

Breaking up the field strength into its radiative and Coulombic components, i.e.
\begin{align}
F_{ur}^{(\pm,d+1)} = F_{ur}^{(R\pm,d+1)} + F_{ur}^{(C\pm,d+1)},
\end{align}
 we can write the charge as
\begin{equation}
\begin{split}\label{chargesep}
\CQ^\pm_\lambda =  \CQ^{\pm S}_\lambda + \CQ^{\pm H}_\lambda,
\end{split}
\end{equation}
 where
\begin{equation}
\begin{split}
\CQ^{\pm S}_\lambda &=  \frac{2}{e^2} \int_{\ci^\pm_\mp} d^d x \,\lambda (1-u\p_u)F_{ur}^{(R\pm,d+1)}\\
\CQ^{\pm H}_\lambda &=  \frac{2}{e^2} \int_{\ci^\pm_\mp} d^d x\, \lambda (1-u\p_u) F_{ur}^{(C\pm,d+1)}  .
\end{split}
\end{equation}
$\CQ^{\pm S}_\l$ are the incoming and outgoing soft charges and $\CQ^{\pm H}_\l$ are the incoming and outgoing hard charges. Using \eqref{cons2} and assuming that the matter current has compact support in $u$, we find that the hard charge can be written as
\begin{equation}
\begin{split}\label{qh1}
\CQ^{\pm H}_\lambda &=  \frac{1}{2d}  \int_{\ci^\pm} du d^d x  \p^a \lambda  \left[ u  \p_a J^{(\pm,d)}_u - (d-1)  J_a^{(\pm,d)} \right] \\
&\qquad \qquad \qquad \qquad \qquad \qquad \qquad \qquad \qquad  +  \frac{2}{e^2}  \int_{\ci^\pm_\pm} d^d x \lambda (1-u\p_u) F_{ur}^{(C\pm,d+1)} . \\
\end{split}
\end{equation}
The second term above receives contributions only from stable massive particles, and hence vanishes in theories with only stable massless particles. For the soft charges, we turn to \eqref{Furlargerodd1} and \eqref{Furlargereven1}, from which we can extract the coefficient of $|r|^{d+1}$ in odd dimensions to be 
\begin{equation}
\begin{split}
(1-u\p_u) F^{(R\pm,d+1)}_{ur} &=  \frac{\G(d)e}{8\pi^{d+1}  }  (-i)^d   \int d^d y \frac{  \p^a \CO^{(\pm,1)}_a ( y ) -  \p^a \CO^{(\pm,1)\dagger}_a ( y )  }{ \left[  ( x - y  )^2  \right]^d } ,
\end{split}
\end{equation}
and that in even dimensions to be
\begin{equation}
\begin{split}\label{evenexp}
&(1-u\p_u) F^{(R\pm,d+1)}_{ur} \\
&\qquad =  \frac{\pm e}{2 \G(\frac{d}{2}+1) }   \sum_{n=2}^\infty \frac{\G(n+1) \left( \frac{iu}{2}\right)^{1-n} }{(4\pi)^{\frac{d}{2}+1}(n-1)}     ( - \p^2 )^{\frac{d}{2}}  \left[ \p^a \CO^{(\pm,n)}_a ( x )  + (-1)^{n-1} \p^a \CO^{(\pm,n)\dagger}_a ( x )  \right]  \\
&\qquad \qquad \qquad \pm \frac{(-1)^{\frac{d}{2}}  \G(d)e}{2^{3-d}\pi } \int d^d y  \frac{1}{  \left[(x-y)^2\right]^{d} }   \left[ \p^a \CO^{(\pm,1)}_a ( y ) + \p^a \CO^{(\pm,1)\dagger}_a ( y ) \right] \\
&\qquad \qquad \qquad \mp \frac{e\left[   \log \left(   \frac{ |u| }{ 4  |r| }  e^{2\g_E} \right)  -  H_{\frac{d}{2}} - 1  \right] }{2(4\pi)^{\frac{d}{2}+1} \G\left( \frac{d}{2}+1\right)   }   ( - \p^2 )^{\frac{d}{2}} \left[ \p^a\CO^{(\pm,1)}_a ( x ) + \p^a\CO^{(\pm,1)\dagger}_a ( x ) \right] \\
&\qquad \qquad \qquad  \mp \frac{i [ \T(u) - \T(r)  ]   e}{16(4\pi)^{\frac{d}{2}} \G \left( \frac{d}{2} + 1 \right)   }  ( - \p^2 )^{\frac{d}{2}} \left[  \p^a \CO^{(\pm,1)}_a ( x )  -  \p^a \CO^{(\pm,1)\dagger}_a ( x ) \right] .
\end{split}
\end{equation}
To determine these coefficients, we assumed the constraint  
\begin{equation}
\begin{split}
\p^a \CO^{(\pm,0)}_{a} ( x )  = \p^a \CO^{(\pm,0)\dagger}_a ( x ) ,
\end{split}
\end{equation}
which was required in \cite{He:2019abc} to cancel the logarithmic divergence in $F_{ur}^{(R\pm,d)}$ in even dimensions. A similar logarithmic divergence is present in \eqref{evenexp}, and to cancel these divergences, we require\footnote{As with the leading soft operator, \eqref{modesconstraint1} implies that we can no longer think of $\CO^{(\pm,1)\dag}_a(x)$ as a creation operator. Rather it is an operator that shifts the vacuum. Note also that there is a relative sign in the constraint, which is consistent with the requirement that the $S$-matrix should be continuous across $\o = 0$.}
\begin{equation}
\begin{split}\label{modesconstraint1}
\p^a \CO^{(\pm,1)}_a ( x )  = -  \p^a \CO^{(\pm,1)\dagger}_a ( x )  . 
\end{split}
\end{equation}
Moreover, this constraint also implies $\left. F_{ur}^{(R\pm,d+1)} \right|_{\ci^\pm_\pm} = 0$ in even dimensions, which was previously assumed to be true in \cite{Kapec:2014zla}. With this, we find that the soft charge takes the form
\begin{equation}
\begin{split}\label{subleadingsoftcharge1}
\CQ^{\pm S}_\lambda &= \begin{cases}
\frac{    i }{2(4\pi)^{\frac{d}{2}} \G \left( \frac{d}{2} + 1 \right)  e } \displaystyle\int d^d z\,   \p^a \CO^{(\pm,1)}_a (z ) ( - \p^2 )^{\frac{d}{2}}  \lambda  (z)  , & d \in 2 \mzz  \\
\frac{ (-1)^{\frac{d+1}{2}}\G(d) i}{2\pi^{d+1} e } \displaystyle\int d^d y \, \p^a \CO^{(\pm,1)}_a ( y )  \displaystyle\int d^d z \frac{ \lambda(z)   }{  \left[  ( z - y  )^2  \right]^d }  , & d \in 2 \mzz + 1 .
\end{cases} \\
\end{split}
\end{equation}

We conclude this section by bringing the soft charge into the same form for both odd and even dimensions by judiciously choosing $\lambda$ to be
\begin{equation}
\begin{split}
\lambda(z) = h_x(z) = \log \left[ ( x - z)^2 \right].
\end{split}
\end{equation}
In even dimensions, we have
\begin{equation}
\begin{split}
( - \p^2 )^{\frac{d}{2}}  h_x  (z) = ( - \p^2 )^{\frac{d}{2}}   \log \left[ ( x - z)^2 \right] = - ( 4\pi )^{\frac{d}{2}} \Gamma \left( \frac{d}{2} \right) \delta^{(d)}(x-z) ,  
\end{split}
\end{equation}
whereas in odd dimensions, we have
\begin{equation}
\begin{split}
 \int d^d z \frac{ h_x(z)  }{ \left[  ( z - y  )^2  \right]^d }  = \frac{ 2 (-1)^{\frac{d-1}{2}} \pi^{d+1} }{ \G ( d + 1  ) }  \delta^{(d)}(x-y) .
\end{split}
\end{equation}
For this choice, we find in all spacetime dimensions that\footnote{Note that one does not lose generality by choosing $\lambda = h_x$, and the matching conditions for $h_x$ imply those for any general function $\l(x)$.}
\begin{equation}
\begin{split}\label{chargegx}
\CQ^{\pm S}_{h_x} &= - \frac{i}{d e }   \p^a \CO^{(\pm,1)}_a ( x )  . 
\end{split}
\end{equation}

\subsection{Ward Identity}

We now turn to semiclassical theories where the quantity of interest is the scattering amplitude
\begin{equation}
\begin{split}
\CA_{n} = \braket{\text{out}}{\text{in}},
\end{split}
\end{equation}
where $n$ is the total number of particles in the scattering amplitude. The classical matching condition \eqref{chargematch} implies the following Ward identity for the amplitude:
\begin{equation}
\begin{split}
	\bra{\text{out}} \left( \CQ^+_\lambda   -   \CQ^-_\lambda \right) \ket{ \text{in} } = 0 . 
\end{split}
\end{equation}
Using \eqref{chargesep}, we can rewrite this as
\begin{equation}
\begin{split}\label{wardid2}
\bra{\text{out}} \left( \CQ^{+S}_\lambda    -    \CQ^{-S}_\lambda \right) \ket{\text{in}} = - \bra{\text{out}}   \left( \CQ^{+H}_\lambda   -   \CQ^{-H}_\lambda  \right) \ket{\text{in}} .
\end{split}
\end{equation}
To simplify this, we now need to determine the action of the charge on one-particle states. Let $\ket{\Psi_i,\vec{p}_i,s_i}$ be a massless one-particle state with charge $Q_i$, momentum $\vec{p}_i$, and spin $s_i$. The massless momentum of this state is parameterized using \eqref{mompar}, so that
\begin{equation}
\begin{split}\label{momstatepar}
p_i^A = \o_i \left( \frac{1+x_i^2}{2} , x_i^a , \frac{1-x_i^2}{2} \right) . 
\end{split}
\end{equation}
The action of the hard charge on bra and ket states is (see Appendix \ref{app:subleading} for an explicit calculation for scalar fields)
\begin{equation}
\begin{split}
\bra{\Psi_i,\vec{p}_i,s_i} \CQ^{+H}_\l&=   \frac{ i   Q_i }{d \o_i  }  \left[  \p^2 \l( x_i  )    \o_i    \p_{\o_i }   - (d-1)  \p^a \l(x_i ) \p_{x_i^a}   \right]   \bra{\Psi_i,\vec{p}_i,s_i}  \\
\CQ^{-H}_\l\ket{\Psi_i,\vec{p}_i,s_i}  &= - \frac{i Q_i  }{d \o_i }  \left[  \p^2 \l( x_i  )  \o_i  \p_{\o_i }    - (d-1) \p^a \l(x_i )   \p_{x_i^a} \right]  \ket{\Psi_i,\vec{p}_i,s_i}  . 
\end{split}
\end{equation}
Using this, we find
 \begin{equation}
 \begin{split}
\bra{\text{out}}  \left( \CQ^{+H}_\lambda   - \CQ^{-H}_\lambda  \right) \ket{\text{in}} =\frac{i}{d} \sum_{i=1}^n \frac{ Q_i }{ \o_i }  \left[  \p^2 \lambda ( x_i )    \o_i \p_{\o_i}   - (d-1)  \p^a \lambda (x_i) \p_{x_i^a}   \right]   \braket{\text{out}}{\text{in}} . 
 \end{split}
 \end{equation}
Setting $\lambda = h_x$ and using \eqref{modesconstraint1} and \eqref{chargegx}, the Ward identity \eqref{wardid2} becomes
\begin{equation}
\begin{split}\label{mainwardid2}
&\bra{\text{out}}\left[ \p^a \CO^{(+,1)}_a ( x )  -      \p^a \CO^{(-,1)}_a ( x ) \right] \ket{\text{in}} \\
&\qquad \qquad \quad = e \sum_{i=1}^n  \frac{ Q_i }{ \o_i }  \left[  \p^2  \log \left[ ( x - x_i)^2 \right]   \o_i \p_{\o_i} + 2 (d-1) \frac{ ( x - x_i )^a }{ ( x - x_i )^2 } \p_{x_i^a}   \right]   \braket{\text{out}}{\text{in}} . 
\end{split}
\end{equation}

\subsection{Soft Theorem}\label{softtheoremchecksubleading}

Finally, we show that the Ward identity \eqref{mainwardid2} is implied by the subleading soft photon theorem. In standard momentum space variables, the soft limit of an amplitude involving an outgoing photon of momentum $\ve_a$ has the universal form
\begin{equation}
\begin{split}\label{outsoftth}
\lim_{p_\g \to 0} \CA^{\text{out}}_{n+1}(\vec{p}_\g,\ve_a ; p_1 , \cdots , p_n ) &= \left[ S_a^\0 + S_a^\1 \right]  \CA_n ( p_1 , \cdots , p_n  ) + O \left( \big(p_\g^0\big)^1 \right) , 
\end{split}
\end{equation}
where
\begin{equation}
\begin{split}\label{softfactors}
S_a^\0 = e \sum_{i=1}^n \eta_i Q_i \frac{ p_i \cdot \ve_a(p_\g) }{ p_i \cdot p_\g } , \qquad S_a^\1 = - i e \sum_{i=1}^n Q_i \frac{ p_\g^A \ve_a^B ( \vec{p}_\g )  }{ p_i \cdot p_\g } \CJ_{i\,AB}  . 
\end{split}
\end{equation}
Here, $\CJ_{i\,AB}$ is the angular momentum operator, which is the sum of the orbital and spin angular momenta:
\begin{equation}
\begin{split}
\CJ_{i\,AB} = \CL_{i\,AB} + \CS_{i\,AB} = - i \left[ p_{iA} \pd{}{p_i^B} - p_{iB} \pd{}{p_i^A} \right]  + \CS_{i\,AB} . 
\end{split}
\end{equation}
The leading soft factor $S^\0_a$ was the central point of discussion in \cite{He:2019abc}. Here, we turn to the subleading soft factor $S^\1_a$, which using the parameterization \eqref{mompar} takes the form
\begin{equation}
\begin{split}
S_a^\1 = e \sum_{i=1}^n \frac{Q_i}{\o_i} \left[ \p^b \log \left[ ( x - x_i )^2 \right]  \left( \delta_{ab} \o_i \p_{\o_i} - i \CS_{i\,ab} \right) - \CI_a{}^b( x - x_i ) \p_{x_i^b} \right],
\end{split}
\end{equation}
where
\begin{equation}
\begin{split}
\CI_{ab}(x)=\delta_{ab} - \frac{2 x_a x_b }{x^2 } 
\end{split}
\end{equation}
is the conformally invariant tensor.

On the other hand, the left-hand-side of \eqref{outsoftth} corresponds to the insertion of the operator $\CO_a^{(+,1)}(x) -  \CO_a^{(-,1)}(x)$ in the $S$-matrix.\footnote{See Appendix C of \cite{He:2019abc} for details.} The subleading soft photon theorem may then be rewritten as
\begin{equation}
\begin{split}
&\bra{\text{out}} \left( \CO_a^{(+,1)}(x)  - \CO_a^{(-,1)}(x)   \right) \ket{ \text{in} } \\
&\qquad \qquad \qquad =  e \sum_{i=1}^n \frac{Q_i}{\o_i} \left[ \p^b \log \left[ ( x - x_i )^2 \right]  \left( \delta_{ab} \o_i \p_{\o_i} - i \CS_{i\,ab} \right) - \CI_a{}^b( x - x_i ) \p_{x_i^b} \right] .
\end{split}
\end{equation}
Taking a divergence of both sides, we find
\begin{equation}
\begin{split}
&\bra{\text{out}} \left( \p^a \CO_a^{(+,1)}(x)  - \p^a \CO_a^{(-,1)}(x)   \right) \ket{ \text{in} } \\
&\qquad \qquad = e \sum_{i=1}^n \frac{Q_i}{\o_i} \left[ \p^2 \log \left[ ( x - x_i )^2 \right] \o_i \p_{\o_i} + 2 (d-1)\frac{ (x - x_i)^b }{ ( x - x_i )^2 }   \p_{x_i^b} \right] \bra{\text{out}} \CS \ket{\text{in}} ,
\end{split}
\end{equation}
which is precisely the Ward identity \eqref{mainwardid2}. Note that even though the subleading soft factor itself depends on the spins of the hard particles, this contribution drops out of the divergence and therefore the Ward identity itself.

We conclude this section with two remarks: 

\begin{itemize}

\item We have successfully shown that the \emph{outgoing} subleading soft photon theorem implies the Ward identity \eqref{mainwardid2}. As was true in \cite{He:2019abc} for the leading soft photon theorem, we could have equally well chosen to work with the incoming soft theorem, which reads
\begin{equation}
\begin{split}
\lim_{p_\g \to 0} \CA^{\text{in}}_{n+1}(\vec{p}_\g,\ve_a ; p_1 , \cdots , p_n ) &= \left[ -  S_a^\0 + S_a^\1 \right]  \CA_n ( p_1 , \cdots , p_n  ) + O \left( (p_\g^0)^1 \right) . 
\end{split}
\end{equation}
This differs form \eqref{outsoftth} by a relative sign. The sign difference in the outgoing and incoming leading soft factor was discussed in \cite{He:2019abc}. To understand the lack sign difference in the subleading soft factor, recall that the incoming subleading Ward identity/soft-theorem is obtained by an insertion of $\p^a\CO_a^{(-,1)\dagger} - \p^a\CO_a^{(+,1)\dagger}$. Using \eqref{modesconstraint1}, this is equal to $\p^a\CO_a^{(+,1)} - \p^a\CO_a^{(-,1)}$, which is precisely equal to the insertion for the outgoing subleading Ward identity. Thus, we see that the outgoing subleading soft photon theorem along with \eqref{modesconstraint1} implies the incoming subleading soft photon thereom, so the latter is not an independent Ward identity of the theory. 

\item While it is true that the leading soft photon theorem is completely equivalent to its corresponding Ward identity (i.e. one implies the other and vice versa), the same is not true for the subleading soft theorem. For instance, the Ward identity does not care about the spins of the particles involved in the scattering amplitude, whereas the soft theorem \eqref{outsoftth} does! Thus, while the subleading soft theorem implies the Ward identity, the inverse is not true. It may be of interest to understand what, if any, asymptotic symmetries the full subleading soft theorem is equivalent to, and we leave this for future work.

\end{itemize}

\subsection{Divergent Large Gauge Transformations}

We have shown that the subleading soft theorem implies the existence of a charge that commutes with the $S$-matrix operator, i.e. it is a symmetry of the $S$-matrix. We now show, using the covariant phase space formalism \cite{Wald:1999wa}, that the charge is associated to divergent large gauge transformations. 

In \cite{He:2019abc}, the authors considered the covariant quantization of abelian gauge theory and constructed the charge that generates gauge transformations on a Cauchy slice $\Sigma$ to be
\begin{equation}
\begin{split}
Q^\Sigma_\ve = \frac{1}{e^2} \int_{\p \Sigma} \ve \ast F . 
\end{split}
\end{equation}
On $\ci^\pm$, this charge simplifies to
\begin{equation}
\begin{split}
Q^\pm_{\hat \ve} = \pm \frac{2}{e^2} \int_{\ci^\pm_\mp} d^d x\, |r|^d {\hat \ve} F_{ur}^{(\pm)} .
\end{split}
\end{equation}
Recalling that $F_{ur}^{(\pm)}$ admits the expansion 
\begin{equation}
\begin{split}
F_{ur}^{(\pm)} (u,r,x) = \frac{F_{ur}^{(\pm,d)}(u,x) }{|r|^d} + \frac{ F_{ur}^{(\pm,d+1)} (u,x)}{ |r|^{d+1} } + \cdots 
\end{split}
\end{equation}
near $\ci^\pm_\mp$, for gauge transformations of the form ${\hat \ve} (u,r,x) = \ve(x)  + O  \big( |r|^{-1} \big)$, we obtain the finite charge
\begin{equation}
\begin{split}
Q^\pm_{\ve} = \pm \frac{2}{e^2} \int_{\ci^\pm_\mp} d^d x\, \ve F_{ur}^{(\pm,d)} .
\end{split}
\end{equation}
It was shown in \cite{He:2019abc} that these charges are associated to the leading soft photon theorem. 

We now turn to divergent gauge transformations. In particular, consider a gauge transformation of the form
\begin{equation}
\begin{split}
{\hat \ve}(u,r,x) = r \lambda(x) + O \left( |r|^{-1} \right).
\end{split}
\end{equation}
The corresponding charge is
\begin{equation}
\begin{split}
\CQ^\pm_\l =   \frac{2}{e^2} \int_{\ci^\pm_\mp} d^d x \left[  |r|  \lambda  \mp    \frac{1}{2d}  u  \p^2 \lambda \right] F_{ur}^{(\pm,d)}  +  \frac{2}{e^2} \int_{\ci^\pm_\mp} d^d x \,\lambda F_{ur}^{(\pm,d+1)\text{fin}}   .
\end{split}
\end{equation}
where we have used \eqref{udep}.

Note that this charge is formally divergent, which is expected since divergent gauge transformations do not preserve the boundary fall-offs on the gauge and matter fields and are therefore not tangent to the phase space. Despite this, it is possible to write down a finite Ward identity since the contribution of the divergent part of the charge to the Ward identity vanishes by virtue of the leading soft photon theorem. The finite part of the charge, which is identical to \eqref{matchcondsubleading} due to \eqref{finitematch1}, is then the only quantity that contributes non-trivially to the Ward identity and, as we have shown, arises from the subleading soft theorem \cite{Campiglia:2016hvg,Laddha:2017vfh,Hirai:2018ijc}.

\section{Non-Abelian Gauge Theories}
\label{sec:nonabelian}

In this section, we study the leading and subleading soft gluon theorem in $(d+2)$-dimensions, which is the non-abelian generalization of the leading and subleading soft photon theorem, and show that the equivalence between asymptotic symmetries and soft theorems continues to hold in this case. The generalization is relatively straightforward and largely involves adding Lie algebra indices on various fields. Hence, we will often refer the reader to previous equations to illustrate the similarities.

\subsection{Notation and Conventions}

We consider a non-abelian gauge theory with Lie algebra $\mfg$. The gauge field is a matrix-valued one-form $A_\mu = A_\mu^I T_I$, where $T_I$ are the anti-Hermitian generators of $\mfg$ in a representation $R$ with index structure $(A_\mu)^i{}_j$. They satisfy $[T_I,T_J]_\mfg = f_{IJ}{}^K T_K$ and $\text{tr}_R [ T_I  T_J ] = \CT_R \delta_{IJ}$, where $\CT_R$ is known as the index of the representation. Throughout this paper, we will use a representation-independent trace, defined by $\tr{\cdots} = \frac{1}{\CT_R} \text{tr}_R [ \cdots ]$, so that $\tr{T_I T_J} = \delta_{IJ}$. Here, we use the subscript $\mfg$ to distinguish the Lie bracket from the quantum commutator. The field strength is a matrix-valued two-form defined by
\begin{equation}
\begin{split}
F_{\mu\nu} = \p_\mu A_\nu - \p_\mu A_\nu + [ A_\mu , A_\nu ]_\mfg . 
\end{split}
\end{equation}
A matter field in representation $R$ is denoted by $\Psi$ and has an index structure $\Psi^i$. Under gauge transformations, the fields transform as
\begin{equation}
\begin{split}
\Psi ~\xrightarrow{g}~ g \Psi , \qquad A_\mu ~\xrightarrow{g}~ g A_\mu g^{-1}  + g \p_\mu g^{-1} , \qquad F_{\mu\nu}~\xrightarrow{g}~g F_{\mu\nu}  g^{-1}  . 
\end{split}
\end{equation}
Infinitesimal gauge transformations are determined by setting $g = 1 + \ve$, where $\ve = \ve^I T_I$. To linear order in $\ve$,
\begin{equation}
\begin{split}
\delta_\ve \Psi = \ve \Psi  , \qquad \delta_\ve A_\mu = - \CD_\mu \ve  , \qquad \delta_\ve F_{\mu\nu} = - [ F_{\mu\nu}, \ve ]  ,
\end{split}
\end{equation}
where $\CD_\mu \ve =  \p_\mu \ve + [ A_\mu , \ve ]$.

Gauge transformations that vanish on the boundary are redundancies of the theory and can be used to choose a gauge. We will henceforth employ the gauge condition
\begin{equation}
\begin{split}
\label{gaugecond}
A_u = 0 . 
\end{split}
\end{equation}

\subsection{Classical Equations and Asymptotics}

The equations of motion are given by (c.f. \eqref{maxwelleq})
\begin{align}\label{eom}
\begin{split}
	\nabla^\mu F_{\mu\nu} &= g_{\YM}^2 {\tilde J}_\nu, \qquad \tilde J_\nu = J_\nu - \frac{1}{g_\YM^2}\left[ A^\mu, F_{\mu\nu}\right]_\mfg , 
\end{split}
\end{align}
where $g_\YM$ is the coupling constant and $J_\nu$ is the conserved matter current. As in the abelian case, we can separate the field strength into radiative and Coulombic parts, denoted $F^{(R\pm)}_{\mu\nu}$ and $F^{(C\pm)}_{\mu\nu}$, respectively. The asymptotic fall-offs near $\ci^\pm$ of $F_{\mu\nu}$ is precisely the same as the abelian case (see equation (2.9) of \cite{He:2019abc}), and we may use them to determine the fall-off conditions for the gauge field $A_\mu$ in the gauge \eqref{gaugecond}:
\begin{align}
\begin{split}
	A_u^{(R\pm)} &= 0, \qquad\qquad\qquad\qquad\qquad\qquad~~ A_u^{(C\pm)} = 0 \\
	A_r^{(R\pm)} &= O\left(|r|^{-\frac{d}{2}-1}\right) + O\left(|r|^{-d}\right), \qquad A_r^{(C\pm)} = \left(|r|^{-d}\right)  \\
	A_a^{(R\pm)} &= O\left(|r|^{-\frac{d}{2}+1}\right) + O\left(|r|^{-d+1}\right), \quad A_a^{(C\pm)} = O\left(|r|^{-d+1}\right).
\end{split}
\end{align}
The radiative field satisfies the sourceless equations and admits a mode expansion identical to \eqref{gaugefieldmodeexp} with the addition of Lie algebra indices on all operators, i.e. $F_{AB}^{(R\pm)} \to F_{AB}^{I(R\pm)}$ and $\CO^\pmm_a \to \CO^{I\pmm}_a$. The large $|r|$ expansion of the field strength is similarly obtained from \eqref{Furlargerodd1} and \eqref{Furlargereven1}.

The asymptotics of the Coulombic field is obtained by assuming a Taylor expansion and then solving the equations order-by-order in large $|r|$. The constraint equations take a form identical to \eqref{cons1} and \eqref{cons2} with the replacement $J_\mu \to {\tilde J}_\mu$ and $e \to g_\YM$. This yields
\begin{equation}
\begin{split}\label{ymcons1}
2\p_u F_{ur}^{(C\pm,d)} &= g_\YM^2 \tilde J_u^{(\pm,d)} \\
\p^2_u F_{ur}^{(C\pm,d+1)}  &=  \mp \frac{g_\YM^2}{4d} \left[  \p^2 {\tilde J}^{(\pm,d)}_u + (d-1)   \p_u \p^a {\tilde J}_a^{(\pm,d)} - 2d \p^2_u {\tilde J}_r^{(\pm,d+2)}  \right] , \end{split}
\end{equation}
where
\begin{align}
\begin{split}\label{curreff}
{\tilde J}_u^{(\pm,d)} &= J_u^{(\pm,d)} + \frac{1}{g_\YM^2}\left[ A^{a\left(R\pm,\frac{d}{2}-1\right)},F_{ua}^{\left(R\pm,\frac{d}{2}-1\right)}\right]_\mfg \\
{\tilde J}_a^{(\pm,d)} &= J_a^{(\pm,d)} + \frac{1}{g_\YM^2}\left[ A^{b\left(R\pm,\frac{d}{2}-1\right)},F_{ab}^{\left(R\pm,\frac{d}{2}-1\right)}\right]_\mfg .
\end{split}
\end{align}
We remark that the explicit form for ${\tilde J}_r^{(\pm,d+2)}$ will not be needed here. 

\subsection{Charges}

Just like the abelian case, we impose the antipodal matching conditions
\begin{align}
\left. F_{ur}^{(+,d)} \right|_{\ci^+_-} &= -\left. F_{ur}^{(-,d)}\right|_{\ci^-_+}  \label{matchcond1} \\
	\left. ( 1 -  u\p_u ) F_{ur}^{(+,d+1)} \right|_{\ci^+_-}  &= \left. ( 1 -  u\p_u )  F_{ur}^{(-,d+1)} \right|_{\ci^-_+} .\label{matchcond2} 
\end{align}
These are now matrix-valued antipodal matching conditions. We introduce matrix-valued functions $\ve(x)$ and $\l(x)$ of the celestial sphere and define the charges
\begin{align}
\begin{split}
Q_\ve^\pm &= \pm\frac{2}{g_\YM^2}\int_{\ci^\pm_\mp} d^dx\,\tr{\ve F_{ur}^{(\pm,d)}}  \\
\CQ_\l^\pm &= \frac{2}{g_\YM^2}\int_{\ci^\pm_\mp} d^dx\,\tr{\l ( 1 -  u\p_u ) F_{ur}^{(+,d+1)} } .
\end{split}
\end{align}
The matching condition implies $Q_\ve^+ = Q_\ve^-$ and $\CQ_\l^+ = \CQ_\l^-$. As before, we can decompose these charge into soft and hard parts
\begin{align}
\begin{split}\label{softhardcharge}
Q^{\pm S}_\ve &= \pm\frac{2}{g_\YM^2}\int_{\ci^\pm_\mp} d^dx\,\tr{\ve F_{ur}^{(R\pm,d)}} \\
\CQ_\l^{\pm S} &= \frac{2}{g_\YM^2}\int_{\ci^\pm_\mp} d^dx\,\tr{ \l ( 1 -  u\p_u ) F_{ur}^{(R+,d+1)} } \\
Q^{\pm H}_\ve &= \pm\frac{2}{g_\YM^2}\int_{\ci^\pm_\mp} d^dx\,\tr{\ve F_{ur}^{(C\pm,d)}} \\
\CQ_\l^{\pm H} &= \frac{2}{g_\YM^2}\int_{\ci^\pm_\mp} d^dx\,\tr{\l ( 1 -  u\p_u ) F_{ur}^{(C+,d+1)} } .
\end{split}
\end{align}

To determine the soft charges, we extract the coefficients of $|r|^{-d}$ and $|r|^{-d-1}$ from \eqref{Furlargerodd1} and \eqref{Furlargereven1}, and impose the constraints 
\begin{align}
	\p^a\CO_a^{I(\pm,0)}(x) = \p^a\CO_a^{I(\pm,0)\dag}(x), \qquad 	\p^a\CO_a^{I(\pm,1)}(x) = - \p^a\CO_a^{I(\pm,1)\dag}(x)
\end{align}
to cancel the logarithmic divergences. Finally, to bring the even and odd dimensional soft charges into a single form, we choose
\begin{equation}
\begin{split}
\ve(z) = f^I_x(z) = T^I (-\p^2 )\log\left[(x-z)^2\right] , \qquad \l(z) = h^I_x(z) = T^I \log\left[(x-z)^2\right] , \\
\end{split}
\end{equation}
which yields
\begin{align}\label{gluonsoft}
	Q_{f_x}^{I,\pm S} = -\frac{1}{g_\YM} \p^a\CO_a^{I(\pm,0)}(x), \qquad \CQ_{h_x}^{I,\pm S} = -\frac{i}{dg_\YM} \p^a\CO_a^{I(\pm,1)}(x) . 
\end{align}

To determine the hard charge, we write the hard charges in \eqref{softhardcharge} as an integral over all of $\ci^\pm$ and use \eqref{ymcons1}. This gives
\begin{align}
\begin{split}
Q^{\pm H}_\ve &=  - \int_{\ci^\pm} du d^d x\, \tr{\ve  \tilde J^{(\pm,d)}_u} \pm \frac{2}{g_\YM^2} \int_{\ci^\pm_\pm}  d^d x\, \tr{\ve F_{ur}^{(C\pm,d)}}  \\
\CQ^{\pm H}_\l &=  \frac{1}{2d}  \int_{\ci^\pm} du d^d x \, \tr{ \p^a \lambda  \left( u  \p_a {\tilde J}^{(\pm,d)}_u - (d-1) {\tilde J}_a^{(\pm,d)} \right)} \\
&\qquad \qquad \qquad \qquad \qquad \qquad \qquad \qquad  +  \frac{2}{g_\YM^2}  \int_{\ci^\pm_\pm} d^d x\, \tr{ \lambda (1-u\p_u) F_{ur}^{(C\pm,d+1)} } . \\
\end{split}
\end{align}
Assuming no stable massive states, the second term in both of the charges above vanish, and we find
\begin{align}\label{gluonhard1}
Q^{\pm H}_\ve &=  - \int_{\ci^\pm} du d^d x\, \tr{\ve  \tilde J^{(\pm,d)}_u} \\
\CQ^{\pm H}_\l &=  \frac{1}{2d} \int_{\ci^\pm} du d^d x \, \tr{ \p^a \lambda  \left( u  \p_a {\tilde J}^{(\pm,d)}_u - (d-1) {\tilde J}_a^{(\pm,d)} \right)}.\label{gluonhard2}
\end{align}

\subsection{Ward Identity and Soft Gluon Theorem}

As in the abelian case, the Ward identity associated to the charges defined in the previous subsection takes the form
\begin{align}\label{gluonward}
\begin{split}
\bra{\text{out}} \left( Q^{+S}_{\ve}    -    Q^{-S}_\ve \right)\ket{\text{in}} &= - \bra{\text{out}}   \left( Q^{+H}_\ve - Q^{-H}_{\ve}  \right) \ket{\text{in}}  \\
\bra{\text{out}} \left( \CQ^{+S}_{\l}    -    \CQ^{-S}_\l \right)\ket{\text{in}} &= - \bra{\text{out}}   \left( \CQ^{+H}_\l - \CQ^{-H}_\l  \right) \ket{\text{in}} . \\
\end{split}
\end{align}
Consider now the right-hand-side of the Ward identities. Let $\ket{\Psi^i,\vec{p},s}$ be a massless one-particle state that transforms in representation $R$, with momentum $\vec{p}$ and spin $s$. We may parameterize this momentum using \eqref{momstatepar}. Then, the action of the leading hard charge has the form (see Appendix \ref{app:nonabcharge} for an explicit computation for scalar and gauge particles)
\begin{equation}
\begin{split}\label{transform1}
Q^{-H}_\ve \ket{ \Psi^i ,\vec p ,s } &= i \ve^I(x) \left(T_I\right)^i{}_j \ket{ \Psi^j ,\vec p ,s } \\
\bra{ \Psi_i ,\vec p ,s } Q^{+H}_\ve & = i \ve^I(x) \bra{\Psi_j ,\vec p ,s } \left( T_I \right)^j{}_i . \\
\end{split}
\end{equation}

A gluon state lives in the adjoint representation and has the form $\ket{F^I,\vec{p},a}$ ($a$ labels the polarization of the gluon). It therefore transforms as
\begin{equation}
\begin{split}\label{transform2}
Q^{-H}_\ve \ket{ F^I , \vec{p} , a }  &=  i\ve^K(x) \big(T_K^{\adj}\big)^I{}_J \ket{ F^J , \vec{p} , a }  =  i f_{KJ}{}^I \ve^K(x) \ket{ F^J , \vec{p} , a }   \\
\bra{ F_I , \vec{p} , a }   Q^{+H}_\ve &=  i \ve^K (x) \bra{F_J ,\vec p , a } \big( T^\adj_K \big)^J{}_I =  i \ve^K (x) f_{KI}{}^J \bra{F_J ,\vec p ,a }  ,
\end{split}
\end{equation}
where we have used the fact that the matrix elements of the generators in the adjoint representation are given by 
\begin{equation}
\begin{split}
\big( T^\adj_I \big)^J{}_K = f_{IK}{}^J . 
\end{split}
\end{equation}
Using this and setting $\ve(z) = f_x(z)$, we find the Ward identity
\begin{equation}
\begin{split}
\label{nonabwardid1}
&\bra{ \text{out} } \left[ \p^a\CO_a^{I(+,0)}(x) - \p^a\CO_a^{I(-,0)}(x) \right] \ket{  \text{in} } \\
&\qquad  \qquad \qquad \qquad \qquad \qquad = -  i g_\YM \sum_{k=1}^n \eta_k \p^2\log\left[ (x-x_k)^2 \right] \bra{\text{out}}T^I_k \ket{\text{in}} , 
\end{split}
\end{equation}
where we have now generalized to include particles transforming under arbitrary representations $R_k$ of $\mfg$, with $T_k^I$ being the generators in that representation. As before, $\eta_k = +1$ for outgoing particles and $\eta_k = -1$ for incoming particles.

In the same manner as above, we can determine the action of the subleading hard charge on the matter states to be
\begin{equation}
\begin{split}\label{transform3}
\CQ^{-H}_\l \ket{ \Psi^i ,\vec p ,s } &= - \frac{1}{d \omega } \left[ \p^2 \l^I (x) \o \p_\o - (d-1) \p^a \l^I(x) \p_a \right] (T_I)^i{}_j \ket{ \Psi^j , \vec{p},s}  \\
\bra{ \Psi_i ,\vec p ,s } \CQ^{+H}_\l & = \frac{1}{d \omega } \left[ \p^2 \l^I (x) \o \p_\o - (d-1) \p^a \l^I(x) \p_a \right] \bra{ \Psi_j ,\vec p ,s }  (T_I)^j{}_i . 
\end{split}
\end{equation}
Similar formulae also hold for the gluon state with $T_I$ replaced with $T^\adj_I$. The Ward identity \eqref{mainwardid2} then generalizes in the non-abelian case to
\begin{equation}
\begin{split}\label{nonabwardid2}
&\bra{\text{out}}\left[ \p^a \CO^{I(+,1)}_a ( x )  -      \p^a \CO^{I(-,1)}_a ( x ) \right] \ket{\text{in}} \\
&~~= - i g_\YM \sum_k  \frac{ 1 }{ \o_k }  \left[  \p^2  \log \left[ ( x - x_k)^2 \right]   \o_k \p_{\o_k} + 2 (d-1) \frac{ ( x - x_k )^a }{ ( x - x_k )^2 } \p_{x_k^a}   \right]   \bra{\text{out}} T^I_k \ket{\text{in}} . 
\end{split}
\end{equation}

Just as in the abelian case, the Ward identity \eqref{nonabwardid1} is equivalent to the leading soft gluon theorem, and the Ward identity \eqref{nonabwardid2} is a consequence of the subleading soft gluon theorem. Recall that the leading and subleading soft gluon theorem in momentum coordinates take the form
\begin{equation}
\begin{split}
\lim_{p_\g \to 0} \CA^{I,\text{out}}_{n+1}(\vec{p}_\g,\ve_a ; p_1 , \cdots , p_n ) &= \left[ S_a^{I\0} + S_a^{I\1} \right]  \CA_n ( p_1 , \cdots , p_n  ) + O \left( \big(p_\g^0\big)^1 \right) , 
\end{split}
\end{equation}
where
\begin{equation}
\begin{split}\label{softfactors11}
S_a^{I\0} = - i g_\YM \sum_{k=1}^n  \eta_k \frac{ p_k \cdot \ve_a(p_\g) }{ p_k \cdot p_\g } T_k^I , \qquad S_a^\1 = - g_\YM \sum_{k=1}^n\frac{ p_\g^A \ve_a^B ( \vec{p}_\g )  }{ p_k \cdot p_\g }  T_k^I \CJ_{k\,AB}  . 
\end{split}
\end{equation}
As was shown in \S\ref{softtheoremchecksubleading}, the kinematic factors above reproduce the kinematic factors in the abelian case \eqref{nonabwardid1} and \eqref{nonabwardid2}. The extra Lie algebra factors are also matched by noting that both the non-abelian Ward identities and soft theorems are obtained from the abelian ones by the replacement $Q_k \to - i T^I_k$ and $e \to g_\YM$. This establishes the relationship between Ward identities and the leading and subleading soft theorems in non-abelian gauge theories.

\section*{Acknowledgements} 
We would like to thank Daniel Kapec and Alok Laddha for useful conversations. TH is grateful to be supported by U.S. Department of Energy grant DE-SC0009999 and by funds from the University of California. PM gratefully acknowledges support from U.S. Department of Energy grant DE-SC0009988.

\appendix

\section{Action of Hard Charges}
\label{hardapp}

In this appendix, we determine the action of the hard charges $Q^{\pm H}_\ve$ and $\CQ^{\pm H}_\l$ on massless one-particle states, both in abelian and non-abelian gauge theories. For simplicity, we focus only on massless scalar and gluon states, but the results of this section are true more generally.

\subsection{Abelian Charges}\label{app:subleading}

Here, we determine the action of the subleading hard charge on a minimally coupled scalar field $\Phi$ with charge $Q$. The corresponding conserved current is 
\begin{equation}
\begin{split}
J_{\mu} = i  Q \left( \Phi ^* \CD_\mu \Phi  - ( \CD_\mu  \Phi  )^* \Phi  \right) , \qquad \CD_\mu \Phi  = \p_\mu \Phi  - i Q  A_\mu \Phi  . 
\end{split}
\end{equation}
The mode expansion for the outgoing/incoming scalar field is
\begin{equation}
\begin{split} \label{scalarfieldexp}
\Phi ^\pmm(X) =   \int \frac{d^{d+1} q}{ ( 2\pi )^{d+1} } \frac{1}{2q^0} \left[  \CO_\Phi^\pmm(\vec{q}\,) e^{ i q \cdot X }  + \CO_{ {\overline \Phi} }^{\pmm\dagger} (\vec{q}\,) e^{ - i q \cdot X }  \right] ,
\end{split}
\end{equation}
where
\begin{equation}
\begin{split}\label{modecomm}
\left[ \CO_\Phi^\pmm (\vec{q}\,) , \CO_\Phi^{\pmm\dagger} (\vec{q}\,') \right] = \left[ \CO_{\overline\Phi}^\pmm (\vec{q}\,) , \CO_{\overline\Phi}^{\pmm\dagger} (\vec{q}\,') \right]  = \big( 2 q^{0}  \big)  (2\pi)^{d+1}  \delta^{(d+1)} \left( \vec{q} - \vec{q}\,' \right) .
\end{split}
\end{equation}
We parameterize the integration variable using \eqref{mompar}. In these variables, we obtain
\begin{equation}
\begin{split} 
\Phi ^\pmm(u,r,x) &=  \frac{1}{2(2\pi)^{d+1}}  \int d^d y  \int_0^\infty d\w \,\o^{d-1} \left[  \CO_{\Phi}^\pmm(\o,x+y) e^{- \frac{i}{2} \o u - \frac{i}{2} \o r y^2  } \right. \\
&\left. \qquad \qquad \qquad \qquad \qquad \qquad \qquad \qquad \qquad + \CO_{\overline\Phi}^{\pmm\dagger} (\o,x+y) e^{ \frac{i}{2} \o u + \frac{i}{2} \o r y^2  }  \right] ,
\end{split}
\end{equation}
where
\begin{equation}
\begin{split}\label{scalarcommrel}
\left[ \CO_{\Phi}^\pmm (\w,x) , \CO_{\Phi}^{\pmm\dagger} (\o',x') \right] &= \left[ \CO_{\overline\Phi}^\pmm (\o,x) , \CO_{\overline\Phi}^{\pmm\dagger} (\o',x') \right] \\
&= 2 \o^{1-d} (2\pi)^{d+1} \delta(\o-\o') \delta^{(d)} ( x - x'  ) .
\end{split}
\end{equation}
The leading order term in the large $r$ expansion of the scalar field is
\begin{equation}
\begin{split} 
\Phi ^\pmm(u,r,x) &=  \frac{\pi}{(2\pi)^{\frac{d}{2}+2}} \int_0^\infty d\w\, \o^{\frac{d}{2}-1}  \left[ \frac{e^{- \frac{i}{2} \o u }}{(ir)^{\frac{d}{2}}}  \CO_{\Phi}^\pmm(\o,x)   + \frac{e^{ \frac{i}{2} \o u   }}{(-ir)^{\frac{d}{2}}} \CO_{\overline\Phi}^{\pmm\dagger} (\o,x)   \right] +  \cdots .
\end{split}
\end{equation}

\subsubsection{Leading Hard Charge}

We begin by studying the leading hard charge $Q^{\pm H}_\ve$. Although this has been worked out in Appendix B of \cite{He:2019abc}, we reproduce here as it will allow us to generalize to the non-abelian case more easily. First, we define the current via normal-ordering to find
\begin{equation}
\begin{split}\label{scalarcurr}
& \int du\, J_u^{(\pm,d)} =  \frac{Q}{2(2\pi)^{d+1} }    \int_0^\infty d\w\, \o^{d-1} \left[  \CO_\Phi^{\pmm\dagger}(\o,x)  \CO_\Phi^\pmm(\o,x)   - \CO_{\overline\Phi}^{\pmm\dagger} (\o,x) \CO_{\overline\Phi}^{\pmm} (\o,x) \right]  . 
\end{split}
\end{equation}
An outgoing or incoming scalar state with charge $Q $ is defined as
\begin{equation}
\begin{split}
\bra{\Phi ,\o  , x  } = \bra{0} \CO_{\Phi}^{\+}(\o ,x )   , \qquad \ket{\Phi ,\o  , x  } =  \CO_{\Phi}^{\-\dagger}(\o ,x )  \ket{0} . 
\end{split}
\end{equation}
Using these definitions, we find
\begin{equation}
\begin{split}
\bra{\Phi ,\o_i  , x_i  } \int du\, J_u^{(+,d)} (u,x) &= Q   \delta^{(d)}(x - x_i ) \bra{\Phi ,\o_i  , x_{i}  }   \\
  \int du\, J_u^{(-,d)} \ket{\Phi ,\o_i  , x_i  }  &=    Q   \delta^{(d)}(x - x_i  ) \ket{\Phi ,\o_i  , x_{i}  } . 
\end{split}
\end{equation}
The action of the leading hard charge is then
\begin{equation}
\begin{split}
\bra{\Phi ,\o  , x  } Q^{+H}_\ve = - Q  \ve(x )\bra{\Phi ,\o  , x  } , \qquad Q^{-H}_\ve \ket{\Phi ,\o  , x  }   = - Q  \ve(x )\ket{\Phi ,\o  , x  }  .
\end{split}
\end{equation}

\subsubsection{Subleading Hard Charge}

As before, we define the current via normal ordering and find
\begin{equation}
\begin{split}\label{currentformulae}
\int du \,u  J^{(\pm,d)}_u &= \frac{-  i  Q }{2(2\pi)^{d+1}}  \int_0^\infty d\o \,\o^{d-1} \left[ \CO_\Phi^{\pmm\dagger}    \p_\o \CO_\Phi^\pmm      -   \p_\o \CO_\Phi^{\pmm\dagger}  \CO_\Phi^{\pmm}    \right. \\
&\left. \qquad \qquad \qquad \qquad \qquad \qquad \qquad \qquad -  \CO_{\overline\Phi}^{\pmm\dagger}     \p_\o   \CO_{\overline\Phi}^{\pmm}    +    \p_\o  \CO_{\overline\Phi}^{\pmm\dagger}    \CO_{\overline\Phi}^{\pmm}     \right]   \\
\int du \,J_a^{(\pm,d)} &= \frac{ i   Q}{2(2\pi)^{d+1}}   \int_0^\infty d\o \,  \o^{d-2} \left[  \CO_\Phi^{\pmm\dagger}  \p_a  \CO_\Phi^\pmm    - \p_a  \CO_\Phi^{\pmm\dagger}   \CO_\Phi^{\pmm}    \right. \\
&\left. \qquad \qquad \qquad \qquad \qquad \qquad \qquad \qquad +  \p_a \CO_{\overline\Phi}^{\pmm\dagger}    \CO_{\overline\Phi}^{\pmm}    - \CO_{\overline\Phi}^{\pmm\dagger}     \p_a \CO_{\overline\Phi}^{\pmm}      \right]  ,
\end{split}
\end{equation}
where we dropped the $(\o,x)$ dependence of the mode coefficients for clarity. It follows
\begin{equation}
\begin{split}
\bra{\Phi,\o_i,x_i} \int du \,u  J^{(+,d)}_u(u,x)  &=  -  i   Q \delta^{(d)}(x-x_i)  \left[ 2 \p_{\o_i}  +  (d-1) \o_i^{-1}  \right] \bra{\Phi,\o_i,x_i}  \\
\bra{\Phi,\o_i,x_i} \int du \,J^{(+,d)}_a(u,x)  &=  i Q  \o_i^{-1} \left[  \delta^{(d)} ( x - x_i ) \p_a     - \p_a \delta^{(d)} ( x - x_i )  \right]   \bra{\Phi,\o_i,x} \\
\int du \, u  J^{(-,d)}_u  \ket{\Phi,\o_i,x_i} &=  i   Q \delta^{(d)} ( x - x_i )   \left[2  \p_{\o_i}   +   ( d - 1 ) \o_i^{-1}  \right]   \ket{\Phi,\o_i,x_i}   \\
\int du \, J_a^{(-,d)}  \ket{\Phi,\o_i,x_i} &=  - Q   i  \o_i^{-1}  \left[  \delta^{(d)} ( x - x_i )  \p_a - \p_a \delta^{(d)} ( x - x_i ) \right]  \ket{\Phi,\o_i,x}  . 
\end{split}
\end{equation}
The action of the hard charge on the states is therefore
\begin{equation}
\begin{split}
\bra{\Phi,\o ,x } \CQ^{+H}_\l &=   \frac{ i   Q }{d \o  }  \left[  \p^2 \l ( x  )    \o    \p_{\o }   - (d-1)  \p^a \l (x ) \p_a   \right]   \bra{\Phi,\o ,x }  , \\
\CQ^{-H}_\l \ket{\Phi,\o ,x }  &= - \frac{i Q }{d\o }  \left[  \p^2 \l (x  )  \o  \p_{\o }    - (d-1) \p^a \l (x )   \p_a \right]  \ket{\Phi,\o ,x }  . 
\end{split}
\end{equation}

\subsection{Non-Abelian Generalization}
\label{app:nonabcharge}

Lastly, we now generalize the results of the previous subsection to non-abelian gauge theories. We consider a minimally coupled scalar field $\Phi^i$ in a representation $R$. The corresponding matter current is
\begin{equation}
\begin{split}
J^I_\mu = \left( \CD_\mu \Phi \right)^\dagger T^I \Phi  - \Phi^\dagger T^I \CD_\mu \Phi , \qquad  \CD_\mu \Phi = \p_\mu \Phi + A_\mu \Phi . 
\end{split}
\end{equation}
The scalar field has a mode expansion identical to \eqref{scalarfieldexp} with the replacement $\CO_\Phi^{(\pm)} \to \CO_\Phi^{i(\pm)}$. The commutators \eqref{scalarcommrel} are also modified by adding an extra factor of $\delta^i_j$ to the right-hand-side, i.e.
\begin{equation}
\begin{split}\label{nonabscalarcommrel}
\left[ \CO_{\Phi}^{i\pmm} (\o,x) , \CO_{j\Phi}^{\pmm\dagger} (\o',x') \right] &= \left[ \CO_{j \overline\Phi}^{\pmm} (\o,x) , \CO_{\overline\Phi}^{i\pmm\dagger} (\o',x') \right] \\
&= 2 \o^{1-d} (2\pi)^{d+1} \delta^i_j \delta(\o-\o') \delta^{(d)} ( x - x'  ) .
\end{split}
\end{equation}
Next, \eqref{scalarcurr} and \eqref{currentformulae} get modified to
\begin{equation}
\begin{split}
\int du \, J_u^{I(\pm,d)} &= \frac{ i (T^I)^i{}_j}{2(2\pi)^{d+1} }    \int_0^\infty d\o \,\o^{d-1} \left[  \CO_{i\Phi}^{\pmm\dagger}(\o,x)  \CO_\Phi^{j\pmm}(\o,x)  \right. \\
&\left. \qquad \qquad \qquad \qquad \qquad \qquad \qquad \qquad \qquad   - \CO_{\overline\Phi}^{j\pmm\dagger} (\o,x) \CO_{i\overline\Phi}^{\pmm} (\o,x) \right]  \\
\int du\, u J^{I(\pm,d)}_u &= \frac{(T^I)^i{}_j }{2(2\pi)^{d+1}}  \int_0^\infty d\o\, \o^{d-1} \left[ \CO_{i\Phi}^{\pmm\dagger}    \p_\o \CO_\Phi^{j\pmm}      -   \p_\o \CO_{i\Phi}^{\pmm\dagger}  \CO_\Phi^{j\pmm}    \right. \\
&\left. \qquad \qquad \qquad \qquad \qquad  \qquad \qquad -  \CO_{\overline\Phi}^{j\pmm\dagger}     \p_\o \CO_{i\overline\Phi}^{\pmm}    +    \p_\o  \CO_{\overline\Phi}^{j\pmm\dagger}    \CO_{i\overline\Phi}^{\pmm}     \right]   \\
\int du\, J_a^{I(\pm,d)} &= \frac{- (T^I)^i{}_j}{2(2\pi)^{d+1}}   \int_0^\infty d\o   \,\o^{d-2} \left[  \CO_{i\Phi}^{\pmm\dagger}  \p_a  \CO_\Phi^{j\pmm}    - \p_a  \CO_{i\Phi}^{\pmm\dagger}   \CO_\Phi^{j\pmm}    \right. \\
&\left. \qquad \qquad \qquad \qquad \qquad \qquad \qquad \quad +  \p_a \CO_{\overline\Phi}^{j\pmm\dagger}    \CO_{i\overline\Phi}^{\pmm}    - \CO_{\overline\Phi}^{j\pmm\dagger}     \p_a \CO_{i\overline\Phi}^{\pmm}      \right]  .
\end{split}
\end{equation}
The $in$ and $out$ states with momentum $\vec{p}$ parameterized by $\o$ and $x$ are defined as 
\begin{equation}
\begin{split}
 \bra{\Phi_i , \o , x } = \bra{0}\CO_{i{\overline \Phi}}^{(+)} (\o,x) , \qquad \ket{\Phi^i , \o , x } = \CO_{\overline\Phi}^{i(-)\dagger} (\o,x)\ket{0}   . 
\end{split}
\end{equation}
Using the definition of the hard charges \eqref{gluonhard1} and \eqref{gluonhard2} and the commutators \eqref{nonabscalarcommrel}, we find 
\begin{equation}
\begin{split}
Q^{-H}_\ve \ket{\Phi^i , \o , x } &=  i \ve^I(x)  (T_I)^i{}_j \ket{\Phi^j , \o , x }  \\
\bra{\Phi_i ,  \o , x }  Q^{+ H}_\ve &=   i  \ve^I \bra{\Phi_j ,  \o , x }  (T_I)^j{}_i \\
\CQ^{-H}_\l  \ket{\Phi^i , \o , x } &= - \frac{ 1 }{ d \o } \left[ \p^2 \l^I \o \p_\o - (d-1) \p^a \l^I \p_a \right]  (T^I)^i{}_j \ket{\Phi^j , \o , x } \\
\bra{\Phi_i , \o , x } \CQ^{+H}_\l  &= \frac{ 1 }{ d \o } \left[ \p^2 \l^I \o \p_\o - (d-1) \p^a \l^I \p_a \right]  \bra{\Phi_j , \o , x } (T^I)^j{}_i .
\end{split}
\end{equation}

To complete the discussion, we must show that these charges also act on the gluon states in the manner given in \eqref{transform2}. This contribution arises from the extra terms in the \emph{effective current} \eqref{curreff}, so we will focus only on this extra term. Explicitly, this is
\begin{align}
\begin{split}\label{currextra}
\tilde J_u^{I(\pm,d)\text{gauge}} &= \frac{f_{JK}{}^I }{g_\YM^2} A^{aJ \left(R\pm,\frac{d}{2}-1\right)} F_{ua}^{K\left(R\pm,\frac{d}{2}-1\right)} \\
{\tilde J}_a^{I(\pm,d)\text{gauge}} &= \frac{f_{JK}{}^I }{g_\YM^2} A^{bJ\left(R\pm,\frac{d}{2}-1\right)} F_{ab}^{K\left(R\pm,\frac{d}{2}-1\right)} .
\end{split}
\end{align}
To determine these currents, we need to determine the large $|r|$ expansion of the gauge field. First, we recall the mode expansion of the gauge field to be
\begin{align}
\begin{split}
A_A^{(R\pm)}(X) &=   g_\YM\int \frac{d^{d+1} q}{ ( 2\pi )^{d+1} } \frac{1}{2q^0} \left[  \ve_{A}^a(\vec q\,)\CO_{a}^\pmm(\vec{q}\,)  e^{ i q \cdot X }  + \ve_A^a(\vec q\,)^* \CO_{a}^{\pmm\dagger} (\vec{q}\,)  e^{ - i q \cdot X }  \right].
\end{split}
\end{align}
All the relevant quantities above are defined after equation \eqref{gaugefieldmodeexp}, and the mode coefficients satisfy
\begin{equation}
\begin{split}
\left[ \CO_{a}^{I\pmm}(\vec{q}\,)  , \CO_b^{J\pmm\dagger}(\vec{q}\,')  \right] = \big(2q^0\big) (2\pi)^{d+1} \delta_{ab} \delta^{IJ} \delta^{(d+1)} ( \vec{q} - \vec{q}\,' ) . 
\end{split}
\end{equation}
Moving to flat null coordinates and using \eqref{mompar}, we find
\begin{align}
\begin{split}
A^{(R\pm)}_a(u,r,x) &=  \frac{g_\YM r}{2(2\pi)^{d+1}} \int  d^d y d\w \,\w^{d-1} \left[ \CO_{a}^\pmm(\w,x+y)  e^{-\frac{i}{2}\w u - \frac{i}{2}\w ry^2}    \right. \\
& \left. \qquad \qquad \qquad \qquad \qquad \qquad \qquad \qquad + \CO_{a}^{\pmm\dag}(\w,x+y)  e^{\frac{i}{2}\w u + \frac{i}{2}\w ry^2}  \right].
\end{split}
\end{align}
The leading order term in the large $r$ expansion of the field is
\begin{equation}
\begin{split}
A^{(R\pm)}_a &=  \frac{- i g_\YM }{2(2\pi)^{\frac{d}{2}+1}} \int  d\w \w^{\frac{d}{2}-1} \left[  \frac{e^{-\frac{i}{2}\w u }  }{ (i r)^{\frac{d}{2}-1} } \CO_{a}^\pmm(\w,x) - \frac{e^{\frac{i}{2}\w u }}{ (-i r)^{\frac{d}{2}-1} }  \CO_{a}^{\pmm\dag}(\w,x)    \right] + \cdots .
\end{split}
\end{equation}
Using this, we find
\begin{equation}
\begin{split}
\int du \,\tilde J_u^{I(\pm,d)\text{gauge}} &= - \frac{ i f_{JK}{}^I   }{2(2\pi)^{d+1}}  \int d\w \,\w^{d-1} \CO^{bJ\pmm\dag}(\w,x) \CO_b^{K\pmm}(\w,x)  \\
\int du \,{\tilde J}_a^{I(\pm,d)\text{gauge}} &= \frac{f_{JK}{}^I }{(2\pi)^{d+1} }  \int  d\w\, \w^{d-2}  \left[   \p_{[a} \CO_{b]}^{K\pmm\dag}(\w',x)   \CO^{aJ\pmm}(\w,x)  \right. \\
&\left. \qquad \qquad \qquad \qquad \qquad \qquad \qquad +  \CO^{aJ\pmm\dag}(\w,x)  \p_{[a}\CO_{b]}^{K\pmm}(\w',x)  \right] \\
\int du \,u \tilde J_u^{I(\pm,d)\text{gauge}}  &= \frac{f_{JK}{}^I }{2(2\pi)^{d+1} } \int  d\w\,   \w^{d-1} \left[ \p_{\o}  \CO_{a}^{J\pmm\dag}(\w,x) \CO^{aK\pmm}(\w,x) \right. \\
&\left. \qquad \qquad \qquad \qquad \qquad \qquad \qquad - \CO^{aJ\pmm\dag}(\w,x)   \p_{\o}   \CO_{a}^{K\pmm}(\w,x)  \right] ,
\end{split}
\end{equation}
where the current is defined via normal ordering.

Incoming and outgoing gluon states are defined as
\begin{align}
\begin{split}
\bra{ F^I,\w,x,a} = \bra{0} \CO^{I\+}_a(\w,x), \qquad \ket{ F^I,\w,x,a } = \CO^{I\-\dag}_a(\w,x) \ket{0} . 
\end{split}
\end{align}
Using \eqref{gluonhard1} and \eqref{gluonhard2}, we find
\begin{equation}
\begin{split}
\bra{ F_I,\w,x,a} Q^+_\ve &=   i \ve^J(x)  \bra{ F_K ,\w,x,a}  (T^\adj_J)^K{}_I  \\
Q^-_\ve \ket{ F^I , \w , x , a } &= i \ve^J(x) ( T^\adj_J)^I{}_K\ket{ F^K , \w , x , a }  . 
\end{split}
\end{equation}
A similar calculation for the subleading charge implies
\begin{equation}
\begin{split}
\bra{ F_I ,\vec p ,s } \CQ^{+H}_\l & = \frac{1}{d \omega } \left[ \p^2 \l^K (x) \o \p_\o - (d-1) \p^a \l^K (x) \p_a \right] \bra{ F_J ,\vec p ,s }  (T^\adj_K)^J{}_I \\
\CQ^{-H}_\l \ket{ F^I ,\vec p ,s } &= - \frac{1}{d \omega } \left[ \p^2 \l^K (x) \o \p_\o - (d-1) \p^a \l^K (x) \p_a \right] (T^\adj_K)^I{}_J \ket{ F^J , \vec{p},s}  . 
\end{split}
\end{equation}

\bibliography{HM-bib}{}
\bibliographystyle{utphys}

\end{document}